\documentclass[twocolumn,nofootinbib,amsmath,amssymb,aps,prd,balancelastpage,superscriptaddress]{revtex4-1}

\usepackage{color}
\usepackage[active]{srcltx}
\usepackage{amsmath,amsfonts,amssymb,amsthm,amstext,amscd,eucal,srcltx}
\usepackage{epsfig,graphicx,bm}
\usepackage{epstopdf, epsf}
\usepackage{dcolumn}
\usepackage{hyperref}

\newcommand{\be}{\begin{equation}}
\newcommand{\ee}{\end{equation}}

\newcommand{\bse}{\begin{subequations}}
\newcommand{\ese}{\end{subequations}}
\newcommand{\bea}{\begin{eqnarray}}
\newcommand{\eea}{\end{eqnarray}}
\newcommand{\ba}{\begin{array}}
\newcommand{\ea}{\end{array}}
\newcommand{\bc}{\begin{center}}
\newcommand{\ec}{\end{center}}

\begin{document}
\preprint{IPM/P-2012/009}  
\vspace*{3mm}

\title{$\mathcal{H}$olographic $\mathcal{N}$aturalness and $\mathcal{C}$osmological $\mathcal{R}$elaxation}%

\author{Andrea Addazi}
\email{andrea.addazi@lngs.infn.it; andrea.addazi@qq.com}
\affiliation{Center for Theoretical Physics, College of Physics Science and Technology, Sichuan University, 610065 Chengdu, China}
\affiliation{INFN sezione Roma {\it Tor Vergata}, I-00133 Rome, Italy}

\begin{abstract}
\noindent

We rediscuss the main Cosmological Problems as illusions originated from our ignorance of the hidden information holographically stored in {\it vacuo}.
The Cosmological vacuum state is full of a large number of dynamical quantum hairs, dubbed {\it hairons}, which dominate the Cosmological Entropy.
We elaborate on the Cosmological Constant (CC) problem, in both the dynamical and time-constant possibilities.
We show that all dangerous quantum mixings between the CC and the Planck energy scales are exponentially suppressed
as an entropic collective effect of the hairon environment. As a consequence, the dark energy scale is UV insensitive to any planckian corrections. 
On the other hand, the inflation scale is similarly stabilized from any radiative effects.
In the case of the Dark energy, we show the presence of a holographic entropic attractor, favoring a time variation of $\Lambda\rightarrow 0$ in future
rather than a static CC case; i.e. $w>-1$ Dynamical DE is favored over a CC or a $w<-1$ phantom cosmology. 
In both the inflation and dark energy sectors, we elaborate on the Trans-Planckian problem, in relation with the recently proposed 
Trans-Planckian Censorship Conjecture (TCC). We show that the probability for any sub-planckian wavelength modes to survive after inflation is 
completely negligible as a holographic wash-out mechanism. In other words, the hairons provide for a holographic decoherence of the 
transplanckian modes in a holographic scrambling time. 
This avoids the TCC strong bounds on the Inflaton and DE potentials. 

\end{abstract}

\maketitle


{\it Introduction}. Apparently, the inflation $+$ $\Lambda$CDM paradigm can explain most of the observational aspects in the Universe.
The {\it Planck} data on the Cosmic Microwave Background (CMB) are fully compatible with the 
existence of dark energy, cold dark matter and an inflation stage in early Universe \cite{Planck}. 
Then, why are most of theoreticians absolutely unsatisfied with it?

The Dark Energy (DE) as well as the inflation lie on a edge line among
phenomenological satisfaction and theoretical puzzles. The situation, in a nutshell, is as follows.  
The large scale cosmological dynamics is dominated by the DE, sourcing for the late Universe acceleration.
 The DE is thought as an exotic fluid providing for a Cosmological Constant (CC) term within the Einstein equation of fields.
The CC fastidiously remains elusive to any attempts of comprehending its origin, existence and stability. 
First of all, it is really not clear if originated from any quantum field vacuum energy 
or if simulated by any IR modifications of gravity. But this is not the worst of the issues.
In both the possibilities, there is a colossal hierarchy problem between the CC and the Planck energy scale,
of $10^{123}$ orders\footnote{See Ref.\cite{Padmanabhan:2007xy} for a review on the CC problem.}. In principle, this may merely appear as an accident, not more insightful than any 
Eddington's considerations on the proton over the Universe sizes (see Ref.\cite{E}). But the main point is 
that, quantum mechanically, nobody knows how the protection of the Cosmological hierarchy from Planck-CC scale mixings is working.
Another puzzling issue regards the very same existence of a stable de Sitter space-time,
referring on the recent discussions on the string-inspired Swampland Conjecture (SC) 
\cite{Obied:2018sgi,Dvali:2013eja,Dvali:2018fqu}.

On the other hand, in the very early Universe dynamics,
there is a series of unsolved issues regarding the inflation mechanism. 
First of all, there is another hierarchy problem between the inflation and the Planck scale. 
Second, there is the {\it Transplanckian Problem} (TP), leading to recent discussions on the Transplanckian Censorship Conjecture (TCC) \cite{Bedroya:2019snp,Bedroya:2019tba,Mizuno:2019bxy}\footnote{Interesting discussions 
on implications on the Primordial Black Hole production in the early Universe can be found in Refs.\cite{Cai:2019igo,Cai:2019dzj}}. 
The TP regards the possibility that trans-planckian quantum fluctuations, initially larger than the Universe Hubble radius, 
can be stretched during inflation, surviving up to a macroscopic classical regime, as an amplifier effect sourced by the Universe inflation. 
Both the issues seem to be as the two sides of the same coin: a clear separation of the quantum gravity and the inflation domains is not well understood.
As a way out to the TP, the TCC was suggested as a strong bound on both the inflation and dark energy potentials.
The TCC is demanded in order to 
avoid for any amplifications of the sub-Planckian modes from the quantum gravity regime to the Universe today. 

\vspace{0.2cm}

Now, let us arrive to the main point. I think that all these confusions originated from 
ignoring the most important issue in Cosmology, pointed out by {\it Roger Penrose}:
the problem of entropy in the Universe \cite{Penrose}. 
Indeed one may consider the all baryons in the Universe,
around the Eddington number 
$$N_{E}\sim 10^{80}\, , $$
corresponding to an average thermal entropy of the Universe today of 
$$S_{CMB}\sim 10^{88}\, ,$$
related to the
CMB temperature of $T_{CMB}\simeq 2.7\, K$. 
Neverthless, if baryons would be re-organized in a maximally entropic state,
i.e. as a Black hole, the entropy would be much larger as 
\begin{equation}
\label{S}
S\sim 10^{123}\,.
\end{equation}

Eq.1 is exactly of the same order of the hierarchy between the cosmological constant and the Planck scale. 
On the other hand, dS/CFT \cite{Strominger:2001pn,Bousso:2001mw} predicts that the entropy of the Universe would be 
\begin{equation}
\label{Sentropy}
S_{\Lambda}\sim \frac{M_{Pl}^{2}}{\Lambda}\sim 10^{123}\, ,
\end{equation}
related to the Hubble area
as 
\begin{equation}
\label{Slam}
S_{\Lambda}\sim \frac{r_{H}^{2}}{L_{Pl}^{2}}\, , 
\end{equation}
where $r_{H}\sim 1/\sqrt{\Lambda}$ is the Hubble radius. 

If de Sitter space-time contains such a large amount of entropy, 
that cannot be accounted from normal matter, then an intrinsic information is stored in the space-time. 
This is leading us to postulate the existence of a large number of dynamical quantum hairs, avoiding the classical no hair theorem, 
accounting for the Universe memory\footnote{Possible quantum hairs were discussed by many authors as a way-out to the Black Hole information paradox, e.g see Refs.\cite{V1,Coleman:1991ku,Preskill:1992tc,Giddings:1993de,Dvali:2012rt,V2,Strominger:2013jfa,Hawking:2016msc,Dvali:2015rea,Chiang:2020lem,Ellis:1991qn,Ellis:2016atb,Addazi:2017xur}.}. 
Therefore 
$$S_{\Lambda}\sim N \sim 10^{123}\,,$$
where $N$ is the number of dynamical hairs, that we dub {\it hairons}. 
Indeed, the configuration space associated to hairons is much larger than the one 
corresponding to the visible sector of the Universe today:
$$P_{\Lambda}\sim e^{10^{123}}>>P_{m}\sim 10^{10^{88}}\, .$$

The same logic can be applied to the Universe during the inflation epoch.
At that time the hierarchy, once again, amounts to be of the same order of the holographic entropy:
$$S_{inflation}\sim \frac{M_{Pl}^{2}}{H}\sim N \sim 10^{12}, $$
having assumed a Hubble rate around the $10^{13}\, {\rm GeV}$ scale.

\vspace{0.1cm}

In this paper, we suggest that the cosmological stability issues can be solved 
as an effective gravitational-mediated decoherence effect induced by the holographically stored quantum hairs.
In other words, the CC and inflation stabilities, the Swamplands and the Transplanckian problem
can be all re-interpreted and addressed considering the entanglement of a large hidden hairon state with the Inflation and DE wave functions.  
This leads to the emergence of a new paradigm dubbed {\it $\mathcal{H}$olographic $\mathcal{N}$aturalness} ($\mathcal{HN}$)\footnote{See also recent discussions of the $\mathcal{HN}$
in contest of particle physics \cite{Addazi:2020hrs} and topological aspects \cite{Addazi:2020mnm}.}; where
the UV divergences are cured from considering a large amount of holographic soft hairs rather than a UV completion in a Wilsonian sense. 
We will show that the $\mathcal{HN}$ principle provides a new razor criterium that can discriminate among 
stable cosmologies, living in the highest probability configuration phase and safe against quantum instabilities,
from unstable solutions in the {\it Holographic Swamplands}.


\vspace{0.2cm}

{\it Dynamical Dark Energy and Inflation.} Let us consider a cosmological scalar field sourcing the Universe acceleration. 
This field can be, for example, a quintessence scalar or a scalaron of an extended theory of gravity (for example in $f(R)$-gravity)
 \cite{Nojiri:2006ri,Nojiri:2010wj,Capozziello:2011et,Bamba:2012cp}.
In this class of models, there are two possible main sources of instabilities. The first is the vacuum energy instability 
from the particle physics sector, the Standard Model or its extensions, sourced by Feynman bubble diagrams. 
In other words, even if we start from an initial bare CC that it is zero, there is no any effect or symmetry 
preventing its UV mixing with $M_{Pl}$. This is surely a common problem of both dynamical and statical dark energy models. 
The second possible issue arrises from UV divergences in the scalar field propagator.
This is very much the same problematic point destabilizing the Higgs boson in the Standard Model. 
The propagator corrections are more controllable than bubble diagrams,
since, if the quintessence couplings with any other SM fields are zero at pertubative level, 
then the UV divergences may be disentangled by the dynamical dark energy sector. 

Here, we show that all quantum radiative instabilities are suppressed by the large entropic contribution from the thermal vacuum. 
Starting from the vacuum energy corrections from quantum bubbles, they correspond to correlators 
as 
\begin{equation}
\label{core}
\langle F(x) F(x)\rangle \equiv \langle F^2\rangle\, ,
\end{equation}
where $F$ corresponds to any possible SM fields or also the same quintessential field and $x$ is a generic space-time coordinate. 
The $\langle ... \rangle$ represents the evaluation of the bubble correlator in the vacuum state. 
Computations performed in QFTs are done considering the vacuum state as a trivial $|0\rangle$. 
This assumption is in contrast with the holographic principle as well as with thermodynamical laws of the space-time.
The vacuum state has a large entropic content, provided by a large number of hidden degrees of freedom, the hairons. 
Therefore, Eq.\ref{core} has to be evaluated on a state that is full of N-hairons, where $N\sim S$ and $S$ is the vacuum entropy:
\begin{equation}
\label{core}
\langle N|F^{2}|N\rangle\, . 
\end{equation}
Now, any bubble diagrams evaluated in the $\langle 0| F^2 |0\rangle$ and UV divergent are interconnected to the 
Eq.\ref{core} through the $\langle N|0\rangle$ and $\langle 0|N\rangle$ amplitudes. 
These are physically interpreted as probability amplitudes to transit to a state with zero entropy to another with N-qubits. 
Clearly, such a transition is exponentially disfavored as 
\begin{equation}
\label{ampla}
\langle N|0\rangle=(\langle 0|N\rangle)^{*}=e^{-N}\sim e^{-S}\, .
\end{equation}
So that, we are ready to evaluate Eq.\ref{core} in the fully entropic space-time: 
it is the same computation in the zero-entropic standard QFT case 
interpolated by the Eq.\ref{ampla} transitions:
\begin{equation}
\label{amm}
\langle N|F^2|N\rangle =\langle N|0\rangle \langle 0| F^{2} |0\rangle \langle 0|N\rangle=e^{-2N}\langle 0| F^{2} |0\rangle\, . 
\end{equation}

This argument is presuming an important fact. The quintessence field wave function has to be fully entangled with the 
hairon state. Assuming that hairons and the DE field are only gravitationally coupled, 
then the scrambling time for a full entanglement follows the
 $$\tau \sim N l_{Pl}^{2} \, {\rm log}\, \, N$$
  holographic scaling 
 \cite{Hayden:2007cs,Sekino:2008he,Shenker:2013pqa}.
Now, in the Minkowski limit, where CC goes to zero, one would say that $S\sim N\sim M_{Pl}^{2}\Lambda^{-1}\rightarrow \infty$
and $\tau \rightarrow \infty$. 

However, looking more carefully to the problem, the apparent divergence of the entropy would appear suspiciously unphysical. 
Indeed, the Hubble radius is, in the case of the dynamical DE field, provided by a dynamical potential\footnote{A dynamical relaxation mechanism can be
considered in a string-theory holographic set-up as in Ref.\cite{Charmousis:2017rof}.} :
$$\Lambda(t)\simeq H^{2}(t)=\frac{1}{2}\dot{\phi}^{2}+V(\phi)\,$$
\begin{equation}
\label{Naa}
 \rightarrow S(t)\sim N(t)\sim M_{Pl}^{2}/\Lambda(t)\, .
\end{equation}
Therefore, the scrambling time-scale is time-varying, but if the field is still around the CC value (as observed)
then Eq.\ref{Naa} is around $\tau \sim 280\, t_{H}$, with $t_{H}$ as the Hubble time.
The rough $280$ factor is obtain as the ${\rm log}\, S\sim {\rm log} \,N\sim {\rm log} \, 10^{123}\sim 123\, {\rm log}\, 10$. 
This is a two digits of relaxation that is necessary for a dynamical full entanglement. 
One can start from a fully entangled Universe or partially or not but still the fate attractor is towards a full entanglement among the 
hairon state and the SM and quintessence ones. 
Finally, regarding the dynamical dark energy, Eq.\ref{Naa} suggests a powerful and insightful fact: 
the number of hairons dynamically varies in the cosmological time dictated by the dynamical cosmological constant. 
This may lead to the temptation to fully identify the quintessence field with the hairon collective field. 
Another interesting observation is that
if the hairons respect the Null Energy Conditions, then second law of thermodynamics 
would impose to their entropy to increase. This means that the effective dynamical $\Lambda(t)$
should decrease in time. This system has a Cosmological attractor towards $\Lambda\rightarrow 0$ 
in an infinite time scaling as $\tau\sim \Lambda^{-1/2}\,{\rm log} \, M_{Pl}^{2}/\Lambda^{2}$. 

In this sense, I propose that the existence of an accelerating source of the Universe can be interpreted as 
a way out to the entropy divergence obtained in not-accelerating Minkowski case. 
Divergences are never desired in physics and the Universe space-time dynamics may be re-interpreted as 
a way out to the the infinite information problem. Therefore, a non-accelerating stage may never be reached within a finite cosmological time. 

The laws of statistical mechanics applied on Eq.\ref{Naa} seem to probabilistically favor
the possibility of a dynamical dark energy with a cosmological attractor $\Lambda\rightarrow 0$ 
rather than a constant $\Lambda$. This seems to point to the same direction of the Swampland conjecture.
However, no every possible dynamical dark energy scenario is probabilistically favored. 
Indeed, for $\Lambda \rightarrow \infty$ in the future, corresponding to a Big Rip, 
the entropy would flow to zero. This seems to forbid any dynamical DE 
models which tend to increase the value of $\Lambda$ in time \footnote{See Refs. \cite{Nojiri:2005sr,Bamba:2008ut,Capozziello:2005tf,Nojiri:2004pf,Nojiri:2004ip,Capozziello:2009hc,Capozziello:2005tf} 
for a panoramic view on the problem of phantom cosmology and singularities.}.

Within this logic, one can easily understand how also propagators as 
\begin{equation}
\label{alla}
\langle N| F(x)F(y)|N\rangle=e^{-2N} \langle 0| F(x)F(y)|0\rangle\, ,
\end{equation}
also have an exponential suppression of any UV divergences. 
Once again, this has to compete with the entanglement entropy
within the hairon state and the $F$ considered.
The attractor solution flows to the relaxation. 
 
Let us explain the whole collective dynamics. 
First, from propagator instabilities, the $F$ particle,
starting from a mass $m_{0}$, acquires a UV contribution that, 
in the worst of the case, is as the Planck scale, $M_{Pl}$.
This can happen
in a transient time that can be Planckian as well,
in a virtual quantum fluctuation. 
Now let us estimate the characteristic time for the relaxation transient.
A mode would have a de Broglie wave length that is entangled with a number of probed hairons inside the corresponding space-time volume. 
The mode will be scrambled 
inside a space volume $L^{3}=m_{0}^{-3}$ after $\tau\sim m_{0}^{-1}{\rm} \log \, M_{Pl}^{2}m_{0}^{-2}$, 
that compared to the particle characteristic time in the rest frame is 
\begin{equation}
\label{mzez}
\frac{\tau}{m_{0}^{-1}}\simeq 2\, \log \frac{M_{Pl}}{m_{0}}\, . 
\end{equation}
This implies that the hierarchy between the Planck scale and the particle mass is scaling as a logarithm rather 
than quadratically. Within this prospective, the fine-tuning problem among the dynamical dark energy scale and the Planck scale is 
of only a factor of $280$ or so. Another important point is that every UV quantum corrections will be inevitably relaxed down to the original bare mass of the particle. 

On the other hand, all UV quantum loops are replaced by analogous ones evaluated as the thermal field theory rules. 
Indeed, the hairon average energy, on a certain volume scales $\bar{V}$, is $\langle \bar{E}\rangle\sim M_{Pl}/\sqrt{\bar{N}}$, where $\bar{V}=(\langle \bar{E}\rangle)^{-3}$.
Any particles with a bare mass $m_{0}$ will probe a volumetric scale of $\bar{V}=m_{0}^{-3}$. Inside the volume $\bar{V}$,
there is a certain number $\bar{N}$ of hairons, in turn related with an average energy of $\langle \bar{E}\rangle\simeq M_{Pl}/\sqrt{\bar{N}}\simeq m_{0}$.
The hairon fields would contribute to the particle bare mass as thermal-like corrections. The one-loop thermal corrections scale as $\Delta m^{2}\simeq cT^{2}\simeq c m_{0}^{2}$, where $T=\langle \bar{E}\rangle$.
Therefore, these contributions are natural, since they all flow to zero in the limit of the particle bare mass going to zero, accordingly with t'Hooft naturalness argument \cite{Naturalness}. 
In the case of inflation, all previous considerations can be repeated, just changing the hierarchical scales considered 
and the inflaton field can be stabilized with the very same mechanism. Indeed, both inflation and dark energy stabilizations are holographically interpolated. 

\vspace{0.2cm}

{\it The Trasplanckian problem and TCC.} Let us now discuss on the Trasplanckian problem in inflation. 
The key towards a $\mathcal{HN}$ reinterpretation of it is through the concept of effective decoherence.
Indeed, if the planckian state loses its coherence, it will not survive to the classical macroscopic Universe.
However, this would demand for an effective collapse of the initial sub-planckian length mode. 
What (or who) is collapsing it?  In our case, as in a thermal bath, the collapse would emerge as a
full entanglement of the trans-planckian mode with the hairon bath, sourced during inflation.
What is the time scale for the entanglement transient?
The answer is, once gain, the scrambling time, that, now, it is dictated by the Hubble scale during inflation:
\begin{equation}
\label{kal}
\tau_{scr}\sim t_{H}\, \log\,\frac{M_{Pl}^2}{H}\sim \sqrt{N}\, l_{Pl}\, \log \, N\,.
\end{equation}
Intriguingly, Eq.\ref{kal} exactly saturates the TCC bound, conjectured as \cite{Bedroya:2019snp,Bedroya:2019tba}\footnote{See also Ref.\cite{Mizuno:2019bxy} for stronger TCC bounds on the tensor-to-scalar ratio $r$ parameter of inflation.}
\begin{equation}
\label{aafi}
\tau\leq \frac{1}{\sqrt{H}}{\rm log}\, \frac{M_{Pl}}{H}\, , 
\end{equation}
related to the bound 
\begin{equation}
\label{akak}
\frac{a_{f}}{a_{i}}L_{Pl}<\frac{1}{H_{f}}\, . 
\end{equation}

Within the N-hairon portrait, the TCC is not imposing any strong constraint to the interaction potential:
the condition 
$$|\nabla V|/V\geq { const}\,\,\,\, \,\,\,\,\, {\rm is\,not\,necessary\,for\,\, the\,\, TP.}$$ 
The TCC is naturally 
satisfied in $\mathcal{HN}$ and the decoherence of sub-planckian length modes is naturally recast
without any particular restriction on the inflaton potential. 
In other words, the $\mathcal{HN}$ solves the Transplanckian problem invoking the environmental effect of the holographic entropy.  
The probability that the modes, $\Phi_{TP}$, from the Planck scale survive after the inflation time is proportional to 
$$P[\Phi_{TP}(t_{Pl})\rightarrow \Phi_{TP}(t\geq t_{Infl})]$$

\begin{equation}
\label{Pa}
< \big|e^{S(t_{Pl})}/e^{S(t_{H_{Inf}})}\big|^2\simeq e^{-2N_{infl}}\simeq e^{-2\times 10^{12}}\, . 
\end{equation}
Of course, this can be extended up to the Dark energy domain, 
with even much more severe suppressions. 

This is implying that any informations on trans-planckian modes is effectively lost. 
Any transplanckian fluctuation is efficiently dissipated in the holographic hairon gas. 
In this sense, {\it both inflation and the Dark Energy are UV Insensitive to any Tranplanckian modes. }

\vspace{0.2cm}

{\it Application: Entropic Razor on Quintessence and Phantom DE.} 
In the Dynamical DE model, one consider the 
scalar field $\phi$ with an equation of state departing from $w=-1$,
as $p=p(\rho)$; coupled with a barotropic fluid with $p=w_{m}\rho$.
In a Friednmann-Lemaitre-Robertson-Walker (FLRW) metric,
the FRLW equations correspond to 
\begin{equation}
\label{eoo}
\dot{\rho}+3H(\rho+p)=-Q\, ,
\end{equation}
\begin{equation}
\label{eood}
\dot{\rho}_{m}+3H(\rho_{m}+p_{m})=Q\, ,
\end{equation}
\begin{equation}
\label{eoot}
\dot{H}=-\frac{\kappa^{2}}{2}(\rho+p+\rho_{m}+p_{m})\, , 
\end{equation}
where $H=\dot{a}/a$ is the Hubble rate, the dot is the derivative with respect to the cosmological time,
$\kappa^{2}=8\pi G$ and $G$ is the Newton's constant.

It is possible to analyze the stability of critical points 
against cosmological perturbations. 

The dynamical system of equations can be conveniently rewritten 
as 
$$\frac{d\mathcal{X}}{d\mathcal{N}}=-[1+p'(\rho)]\Big[3+\frac{Q}{H(\rho+p)} \Big]\mathcal{X}$$
\begin{equation}
\label{e1}
+3\mathcal{X}[2\mathcal{X}+(1+w_{m})(1-\mathcal{X}-\mathcal{Y})]\, , 
\end{equation}
$$\frac{d\mathcal{Y}}{d\mathcal{N}}=-[1-p'(\rho)]\Big[3\mathcal{X}+\frac{Q}{H(\rho-p)}\mathcal{Y}\big]$$
\begin{equation}
\label{e2}
+3\mathcal{Y}[2\mathcal{X}+(1+w_{m})(1-\mathcal{X}-\mathcal{Y})]\, ,
\end{equation}
\begin{equation}
\label{e3}
\frac{1}{H}\frac{dH}{d\mathcal{N}}=-\Big[3\mathcal{X}+\frac{3}{2}(1+w_{m})(1-\mathcal{X}-\mathcal{Y}) \Big]
\end{equation}
with the constraint
\begin{equation}
\label{Omega}
\Omega_{m}=1-\mathcal{X}-\mathcal{Y}\, .
\end{equation}
In the system above, we defined $\mathcal{N}\equiv log\, a$ and $p'(\rho)\equiv dp/d\rho$.

A simplified solution is obtain assuming $Q=0$, $w_{m}=0$ (dust-like matter and no-collisions)
and assuming $d\mathcal{X}/d\mathcal{N}=d\mathcal{Y}/d\mathcal{N}=0$.
In this case, the system has two attractors,
corresponding to $(\mathcal{X},\mathcal{Y})=(0,0)$, in the case of matter dominance,
and $(\mathcal{X},\mathcal{Y})=(0,1)$ for DE domination. 
The second solution correspond to the equation
\begin{equation}
\label{shsh}
\frac{1}{H}\frac{dH}{dN}=-\frac{3}{2}(1+w)\, .
\end{equation}

Therefore, the Hubble rate will increase for $w<-1$.
This corresponds to the following equation for the entropy:
\begin{equation}
\label{entropy}
\frac{1}{S}\frac{dS}{d\mathcal{N}}=c(1+w)\rightarrow S\sim e^{c(1+w)\mathcal{N}}=a^{c(1+w)}\, . 
\end{equation}

In our picture, the $w<-1$ condition is interpreted as a dynamical annihilation of hairons 
in vacuo, leading to a decreasing of information and entropy. 
On the contrary, $w>-1$ corresponds to a creation of hairon information in the Universe.
Finally, $w=-1$ is just the CC case, that corresponds to a conservation of information in the Universe.
This leads to the probabilistic conclusion that the $w<-1$ seems to be favored. 

The entropic suppression of Radiative mixing with the Planck scale is reduced 
as 
$$e^{-N}=e^{-S}=e^{-a^{c(1+w)}}\, ,$$
that is exponentially sensitive to $w$ departures from $-1$. 

The stability of the system can be studied 
consider perturbations around the fixed points 
as 
\begin{equation}
\label{p1}
\frac{d}{dN}{\bf V}=\mathcal{K}{\bf V}\, ,
\end{equation}
where ${\bf V}=(\delta \mathcal{X},\delta\mathcal{Y})^{T}$
is the perturbation vector, with 
\begin{equation}
\label{XY}
(\mathcal{X},\mathcal{Y})^{T}=(\mathcal{X}_{0},\mathcal{Y}_{0})+{\bf V}\, , 
\end{equation}
and $\mathcal{K}_{ij}$ is a matrix with components 
\begin{equation}
\label{c1}
\mathcal{K}_{11}=3(2\mathcal{X}_{0}-\mathcal{Y}_{0}-w)\, , 
\end{equation}
\begin{equation}
\mathcal{K}_{12}=-3\mathcal{X}_{0}\, ,
\label{c2}
\end{equation}
\begin{equation}
\mathcal{K}_{21}=3(-1+\mathcal{Y}_{0}+w)\, , 
\label{c3}
\end{equation}

\begin{equation}
\mathcal{K}_{22}=3(1+\mathcal{X}_{0}-2\mathcal{Y}_{0})\, . 
\label{c4}
\end{equation}

Eq.\ref{p1} allows to related the stability around the critical points
against cosmological perturbations with the Lyapunov eigenvalues of 
$\mathcal{K}$. In the case of DE domination, the eigenvalues are:
\begin{equation}
\label{aka}
\lambda_{1}=3w,\,\,\,\, \lambda_{2}=3(w+1)\, . 
\end{equation}
This shows that the $w<-1$, corresponding to $e^{-\lambda_{1,2}t}$ modes,
is stable against cosmological perturbation. Therefore, the entropic razor 
does not trivialize to prohibiting a cosmological unstable case; it provides a novel 
no go criterium. 

As an interesting case, we can consider a model transiting from 
a phantom-like dynamics with $w<-1$ to a $w>-1$. 
In this case, the phantom model seems to be disfavored only for a transient.
This is allowed by the holographically entropic razor, as a transient 
out of thermal equilibrium, returning to thermalization. 
If we consider our Cosmological state as a open system entangled with a hidden environment, 
the negative entropy flows provoke an apparent violation of laws of thermodynamics. 
An example when the cross-over between $w<-1$ and $w>-1$ is realized is provided by
\begin{equation}
\label{transition}
a(t)=a_{0}\Big(\frac{t}{\bar{t}-t}\Big)^{n}
\end{equation}
where $n>0$ and $0<t<\bar{t}$. 
The corresponding Hubble rate is 
\begin{equation}
\label{HH}
H=n\Big(\frac{1}{t}+\frac{1}{\bar{t}-t}\Big)^{2}\, , 
\end{equation}
in turn related to an entropy 
\begin{equation}
\label{ente}
S\sim N\sim \frac{t(\bar{t}-t)}{\bar{t}}\, . 
\end{equation}
Around $t\simeq \bar{t}$, the entropy and the number of hairon are going to zero, while the Hubble rate diverges. 
This case corresponds to an effective $w$ as 
\begin{equation}
\label{ww}
w=-1-\frac{2}{3n}<-1\, .
\end{equation}
On the other hand, for $t<<\bar{t}$, the effective $w$ is 
\begin{equation}
\label{aka}
w=-1+\frac{2}{3n}>1\, . 
\end{equation}
Therefore, an apparent $w<-1$ in cosmology may be a hint the our Universe is not an isolated system
and that an information flow is entering in, as a dissipative dynamics. 
On the other hand, assuming that the Universe is isolated,
$\mathcal{HN}$ disfavors phantom cosmology and finite future Big Rip singularities. 

\vspace{0.2cm}

{\it Holographic razor on $f(R)$-gravity}. 
The Holographic razor can be applied on the gravitational sector of extended theories of gravity.
As an example, we will consider $f(R)$-gravity models:
\begin{equation}
\label{Ss}
S=\frac{1}{2\kappa^{2}}\int d^{4}x\, \sqrt{-g}[f(R)+2\kappa^{2}\mathcal{L}_{m}]\, ,
\end{equation}
where $\kappa^{2}=8\pi G$ and $\mathcal{L}_{m}$ is the matter Lagrangian. 
In the extreme case of $w<<-1$, the entropic suppression of quantum instabilties
will be un-efficient, leading cosmology to a purely quantum gravity domain.

The field equations, obtained with the variation of the Eq.\ref{Ss} with respect to the metric 
tensor $g_{\mu\nu}$, read as follows:
\begin{equation}
\label{EoM}
R_{\mu\nu}f_{R}-\frac{1}{2}g_{\mu\nu}f+g_{\mu\nu} \nabla_{\alpha} \nabla^{\alpha} f_{R}-\nabla_{\mu} \nabla_{\nu} f_{R}=\kappa^{2}T_{\mu\nu}^{m}\, . 
\end{equation}
where $f_{R}=df/dR$. 
In the FLRW background, this reduces to 
\begin{equation}
\label{HH}
H^{2}=\frac{1}{3f_{R}}\Big[\kappa^{2}\rho_{m}+\frac{Rf_{R}-f}{2}-3H\dot{R}f_{RR} \Big]\, , 
\end{equation}
$$\mathcal{K}=-3H^{2}-2\dot{H}$$
\begin{equation}
\label{tHH}
\mathcal{K}=\frac{1}{f_{R}}\Big[\kappa^{2}p_{m}+\dot{R}^{2}f_{RRR}+2H\dot{R}f_{RR}+\ddot{R}f_{RR}+\frac{1}{2}(f-Rf_{R})\Big]\, .
\end{equation}

The DE sector can be sourced by the $f(R)=R+F(R)$ sector as 
\begin{equation}
\label{FF}
\Omega_{F(R)}=\frac{1}{3H^{2}}\Big(\frac{RF_{R}-F}{2}-3H\dot{R}F_{RR}-3H^{2}F_{R} \Big)\, . 
\end{equation}

The entropic razor can be applied on Eq.\ref{HH} 
as
\begin{equation}
\label{S}
S=3f_{R}\frac{1}{\kappa^{2}\rho_{m}+\frac{Rf_{R}-f}{2}-3H\dot{R}f_{RR}}\, .
\end{equation}
(we omitted the numerical prefactor on $S$ and $H^{-1}$ and we consider $L_{Pl}=1$ as inessential 
for our following discussions). 
The entropic condition 
\begin{equation}
\frac{dS}{d\mathcal{N}}>0\, , 
\end{equation}
highly restrict the possibility on the viable $f(R)$ models. 

The entropic razor does not allowed for 
$w<-1$, leading to the constraint 
\begin{equation}
\label{wee}
w_{eff}=\frac{p_{F(R)}+p_{m}}{\rho_{F(R)}+\rho_{m}}=-1-\frac{2\dot{H}(t)}{3H^{2}(t)}> -1\, . 
\end{equation}
This bound also implies that 
\begin{equation}
\label{b}
\frac{2(1+z)H'(z)}{3H(z)}<0\, ,
\end{equation}
where $1+z=a_{0}/a(t)$ defined the redshift relation with the scale factor
and 
\begin{equation}
\label{akak}
H^{2}(z)=\frac{1}{3f_{R}}\Big[\kappa^{2}\rho_{m}(z)+\frac{R(z)f_{R}-f}{2}+3(1+z)H^{2}f_{RR}R'(z) \Big]\, .
\end{equation}

Let us consider two possible models of $f(R)$ gravity considered 
as viable candidates for Dynamical DE: The Hu-Sawicki \cite{Hu:2007nk} and the Nojiri-Odintsov \cite{Nojiri:2007cq} models
\begin{equation}
\label{FHS}
F_{HS}(S)=-R_{0}\frac{k_{1}(R/R_{0})^{n}}{k_{2}(R/R_{0})^{n}+1}\, , 
\end{equation}
\begin{equation}
\label{NO}
F_{NO}(R)=\frac{R^{n}(aR^{n}-b)}{1+cR^{n}}\, ,
\end{equation}
where $k_{1},k_{2},n,a, b,c,m$ are real free parameters.
These models are particularly interesting since passing Solar System and cosmological tests.
On the other hand, there are regions of the parametric spaces
that, in both the model, dynamically lead to a decreasing and re-increasing of the entropy content in the Universe.

In the HS case, for $c_{1}=2, c_{2}=1,n=1$, the $w_{eff}(z)$ has an excursion 
to the $w<-1$ within the redshift interval $-1<z<1$.
The same happens in the case of 
 the NO model, for a parametric choice of $n=1,a=0.1/H_{0}^{2},b=1,c=0.05/H_{0}$.
 In these two examples, the two models contradict the entropic bound. 
Indeed, the entropic information is related to the hairon temperature 
that can decrease if and only if our Universe information is not closed
and in a dissipative interaction with another hidden sector. 

\vspace{0.1cm}

{\it Interacting Dark energy.} The Holographic entropic bound brings interesting implications on 
models of Interacting Dark Energy (IDE) recently analyzed, e.g. see Refs. \cite{Costa:2013sva,Abdalla:2014cla,Costa:2018aoy,Li:2019loh,Elizalde:2018ahd,Addazi:2018xoa}.  
These models can be mapped in the system of equations Eqs.\ref{eoo},\ref{eood},\ref{eoot}, with a Q-factor different than zero. 
  
The Q-factor can be expressed as 
$$Q(t)=\frac{1}{\kappa^{2}w_{DE}}\Big[9(1+w_{DE})H^{3}+6(2+w_{DE})H\dot{H}$$
\begin{equation}  
\label{Qt}
+2\ddot{H}-\frac{\dot{w}_{DE}}{w_{DE}}(3H^{2}+2\dot{H})\Big]\, . 
\end{equation}
In general, the solution of a IDE system can be highly non-trivial as a integro-differential problem:
\begin{equation}
\label{intn}
H^{2}-\frac{\kappa^{2}}{3}(\hat{\rho}_{m}+\hat{\rho}_{DE})=\frac{\kappa^{2}}{3}\mathcal{I}_{1}\, , 
\end{equation}  
\begin{equation}
\label{kak}
\mathcal{I}_{1}=\hat{\rho}_{m}\int \frac{Q}{\hat{\rho_{m}}}dt-\hat{\rho}_{DE}\int \frac{Q}{\hat{\rho}_{DE}}dt\, ,
\end{equation}
\begin{equation}  
\label{Ha}
\dot{H}+\frac{\kappa^{2}}{2}[\hat{\rho}_{m}+(1+w_{DE})\hat{\rho}_{DE}]=-\frac{\kappa^{2}}{2}\mathcal{I}_{2}
\end{equation}
\begin{equation}
\label{Saa}
\mathcal{I}_{1}=\hat{\rho}_{m}\int \frac{Q}{\hat{\rho}_{m}}dt-(1+w_{DE})\hat{\rho}_{DE}\int \frac{Q}{\hat{\rho}_{DE}}dt\, . 
\end{equation}
 Here, we introduce $\hat{\rho}_{DE,m}$ as the energy densities at $Q=0$.  
  
The entropic razor is expected to impose bounds on $H$ and $Q$ dynamics.   

A simplified class of cases is obtained 
assuming a contant $w_{DE}$.
In this case, the Hubble factor has a form 
\begin{equation}
\label{Hf}
H(t)=H_{1}+\frac{2}{3t}+H_{2}(\hat{t}-t)^{\gamma}\, ,
\end{equation}
while the entropy is inverse proportional to it.
Here we defined $H_{1,2},\gamma$ as integration constant parameters. 
For $\gamma<0$, 
the Hubble rate has a singularity that 
is leading to a zero entropy. 
Therefore this case is forbidden by the holographic razor. 
This constraint percolates on a bound on the 
$Q(t)$, that is 
\begin{equation}
\label{Qtt}
Entropic\, Razor:\,\,\,Q(t)\simeq (t-\hat{t})^{\gamma-2}\rightarrow \gamma>0\, . 
\end{equation}
In principle, if $0<\gamma<2$, the $Q$-factor can diverge without 
leading to a decreasing of the entropy.
In this case, around the $\hat{t}$, perturbation theory breaks down
and the dynamics would become highly non-trivial.
Indeed if interactions between dark or ordinary matter and DE
will explode, then the entropic suppressions of bubble vacuum diagrams 
as well as propagator loops cannot be controlled anymore although 
also the entropy is diverging around $\hat{t}$. 
Therefore, we arrive to a subtle point of the holographic razor:
there are not only no go from classical thermodynamical effects
but also from quantum dangerous mixings. 
In other words, it is impossible, 
to consider a scenario with an infinite interaction and a stable $H$ and $S$
without any quantum destabilizations.
Quantum effects will mix the DE with the Planck scale and re-drive the system 
to a zero entropy UV domain. 
Therefore, the quantum stability requirement enhances the bound on the Q-factor 
and to the Hubble rate to 

\begin{equation}
\label{Qtt}
Quantum\, Stability:\,\,\, Q(t)\simeq (t-\hat{t})^{\gamma-2}\rightarrow \gamma\geq2\, . \\
\end{equation}
$$ $$

 The quantum bound completely eliminates any possible scenarios  
 predicting for any {\it Big Rip}, {\it Big Freeze} and {\it Sudden} singularities

\vspace{0.2cm}

{\it Conclusions and remarks}. In this paper, we have discussed the implications of the $\mathcal{H}$olographic $\mathcal{N}$aturalness
to the Dark energy and Inflation. The Holographic principle suggests that the Universe is populated by a large number of 
holographic quantum hairs, dubbed hairons. The hairon wave function $\Psi(h_{1},...,h_{N})$ is entangled with the Standard model 
states in a holographic scrambling time $\tau \sim \sqrt{N}t_{Pl}\, {\rm log}\, N$.
This is inducing the effective and gravitationally mediated decoherence effects that provide the solutions to many puzzles of contemporary Cosmology.
First of all, we showed that  the $\mathcal{H}\mathcal{N}$ principle efficiently stabilizes the dark energy from UV quantum corrections.
Indeed, the quantum vacuum energy does not accumulate in space-time as a planckian energy density
since dissipated inside the large vacuum thermality. 
We also argued that $\mathcal{H}\mathcal{N}$ probabilistically favors models with a Dynamical variation of $\Lambda$ towards a cosmological attractor 
$\Lambda\rightarrow 0$ rather than a static CC. On the other hand, the Big Rip, predicted from scenarios with a $\Lambda$ increasing in time,
is entropically sequestered as a quantum unstable state. Indeed $\mathcal{H}\mathcal{N}$ provides a new probabilistic criterium for distinguishing among {\it Holographic Swamplands} (HS) and {\it Holographic Landscapes} (HL).
Then, we have shown that the $\mathcal{HN}$ has several strong implications in quintessence DE, phantom cosmology, $f(R)$-gravity and interacting DE-DM models. 
On the other hand, also the Inflation scale does not mix with the Planckian domain, from the very same mechanisms explored in the case of DE. 
Finally, we moved towards a discussion of the Transplanckian problem, in connection with the Transplanckian Censorship Conjecture (TCC).
The $\mathcal{HN}$ provides a strong confirmation of the TCC but a novel re-interpretation of it:
 the hairon decoherence  efficiently washes out any transplanckian modes in a holographic scrambling time, 
 in turn exactly saturating the TCC bound! Therefore, the TCC does not impose any bounds on the inflaton and the dark energy sectors:
the trans-planckian problem is naturally solved as a holographic collective effect. 
In this sense, inflation and dark energy are {\it UV insensitive} to any transplanckian effects.

\vspace{0.2cm}

{\it Acknowledgements}. We wish to thank {\it S. Pi} for valuable comments on these aspects.

\onecolumngrid


\twocolumngrid

\end{document}